\documentclass[11pt,twoside]{article}
\usepackage{asp2004}
\usepackage{psfig}
\usepackage{epsf}
\usepackage{graphics}
\usepackage{lscape}
\def\edcomment#1{\iffalse\marginpar{\raggedright\sl#1\/}\else\relax\fi}
\reversemarginpar
\begin{document}
\title{Hypernovae, Black-Hole-Forming Supernovae, and First Stars}
\author{K. Nomoto, 
N. Tominaga, 
H. Umeda, 
K. Maeda, 
T. Ohkubo, 
J. Deng, 
P.A. Mazzali
}

\affil{Department of Astronomy and Research Center for the Early
  Universe, University of Tokyo, Bunkyo-ku, Tokyo 113-0033, Japan\\
  INAF-Osservatorio Astronomico, Via Tiepolo, 11,
  34131 Trieste, Italy
}

\begin{abstract}

Recent studies of core-collapse supernovae have revealed the existence
of two distinct classes of massive supernovae (SNe): 1) very energetic
SNe (Hypernovae), whose kinetic energy (KE) exceeds $10^{52}$\,erg,
about 10 times the KE of normal core-collapse SNe, and 2) very faint
and low energy SNe ($E \mathrel{\rlap{\lower 4pt \hbox{\hskip 1pt
$\sim$}}\raise 1pt \hbox {$<$}}$ 0.5 $\times$ $10^{51}$\,erg; Faint
supernovae).  These two classes of supernovae are likely to be
"black-hole-forming" supernovae with rotating or non-rotating black
holes.  We compare their nucleosynthesis yields with the abundances of
extremely metal-poor (EMP) stars to identify the Pop III (or first)
supernovae.  We show that the EMP stars, especially the C-rich type,
are likely to be enriched by black-hole-forming supernovae.

\end{abstract}
\thispagestyle{paspcstitle}

\section{Introduction and Summary}

Stars more massive than $\sim$ 25 $M_\odot$ form a black hole at the
end of their evolution.  Stars with non-rotating black holes are
likely to collapse "quietly" ejecting a small amount of heavy elements
(Faint supernovae).  In contrast, stars with rotating black holes are
likely to give rise to very energetic supernovae (Hypernovae).  We
present distinct nucleosynthesis features of these two classes of
"black-hole-forming" supernovae.  Nucleosynthesis in Hypernovae is
characterized by larger abundance ratios (Zn,Co,V,Ti)/Fe and smaller
(Mn,Cr)/Fe than normal supernovae, which can explain the observed
trend of these ratios in extremely metal-poor stars.  Nucleosynthesis
in Faint supernovae is characterized by a large amount of fall-back.
We show that the abundance pattern of the recently discovered most
Fe-poor star, HE0107-5240, and other extremely metal-poor stars are in
good accord with those of black-hole-forming supernovae, but not
pair-instability supernovae.  This suggests that black-hole-forming
supernovae made important contributions to the early Galactic (and
cosmic) chemical evolution.  Finally we discuss the nature of First
(Pop III) Stars.

\section{Hypernovae and Faint Supernovae}

Type Ic Hypernovae 1998bw and 2003dh were clearly linked to the
Gamma-Ray Bursts GRB 980425 (Galama et al. 1998) and GRB 030329
(Stanek et al. 2003; Hjorth et al. 2003), thus establishing the
connection between long GRBs and core-collapse supernovae (SNe).
SNe~1998bw and 2003dh were exceptional for SNe~Ic: they were as
luminous at peak as a SN~Ia, indicating that they synthesized 0.3 -
0.5 $M_\odot$ of $^{56}$Ni, and their kinetic energy (KE) were
estimated as $E_{51} = E/10^{51}$\,erg $\sim$ 30 (Iwamoto, Mazzali,
Nomoto, et al. 1998; Woosley, Eastman, \& Schmidt 1999; Nakamura et
al. 2001a; Mazzali et al. 2003).

Other ``hypernovae'' have been recognized, such as SN~1997ef (Iwamoto
et al. 2000; Mazzali, Iwamoto, \& Nomoto 2000), SN~1999as (Knop et
al. 1999; Hatano et al. 2001), and SN~2002ap (Mazzali et al. 2002).
These hypernovae span a wide range of properties, although they all
appear to be highly energetic compared to normal core-collapse SNe.
The mass estimates, obtained from fitting the optical light curves and
spectra, place hypernovae at the high-mass end of SN progenitors.

\begin{figure}[t]
\plottwo{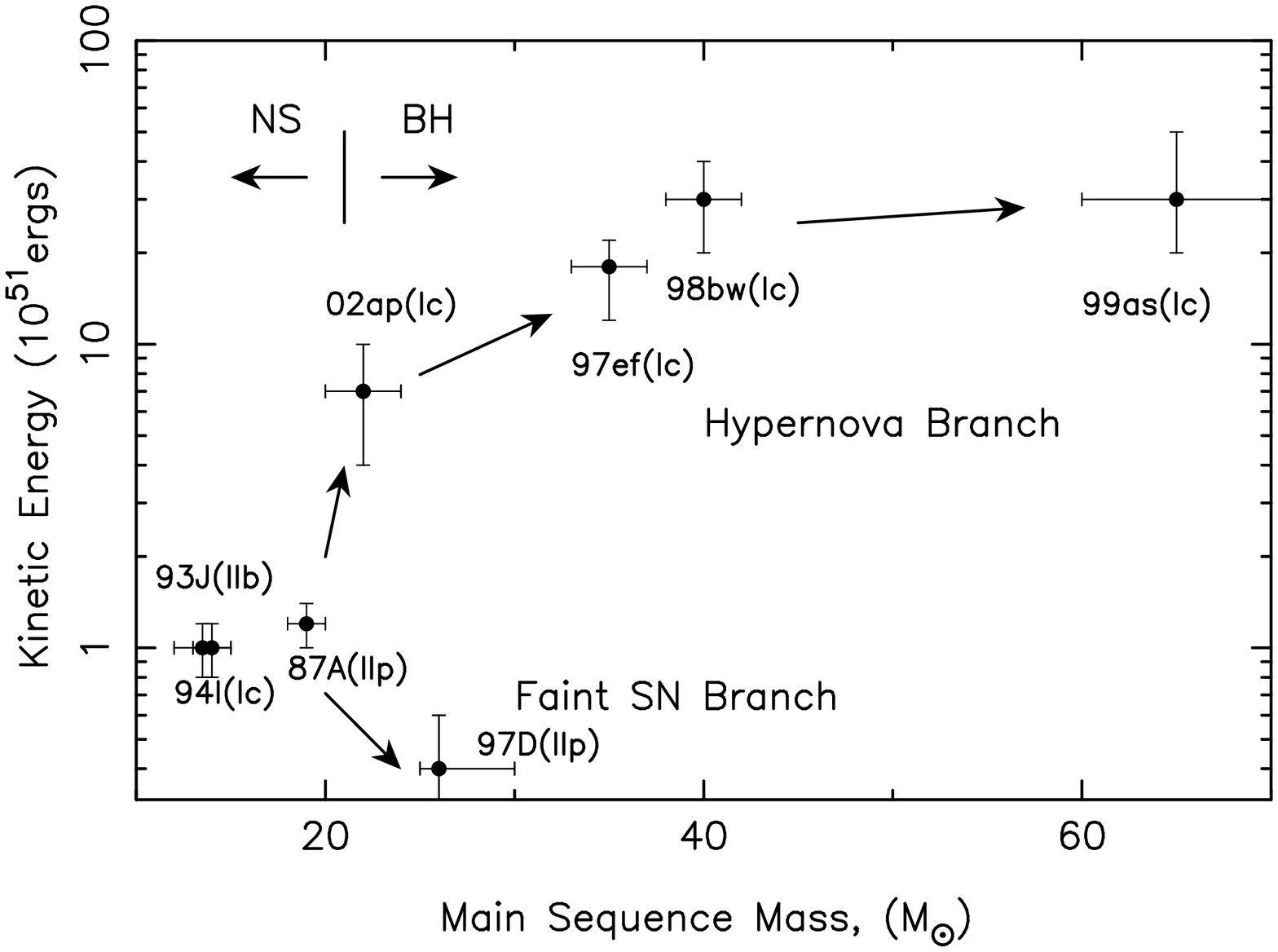}{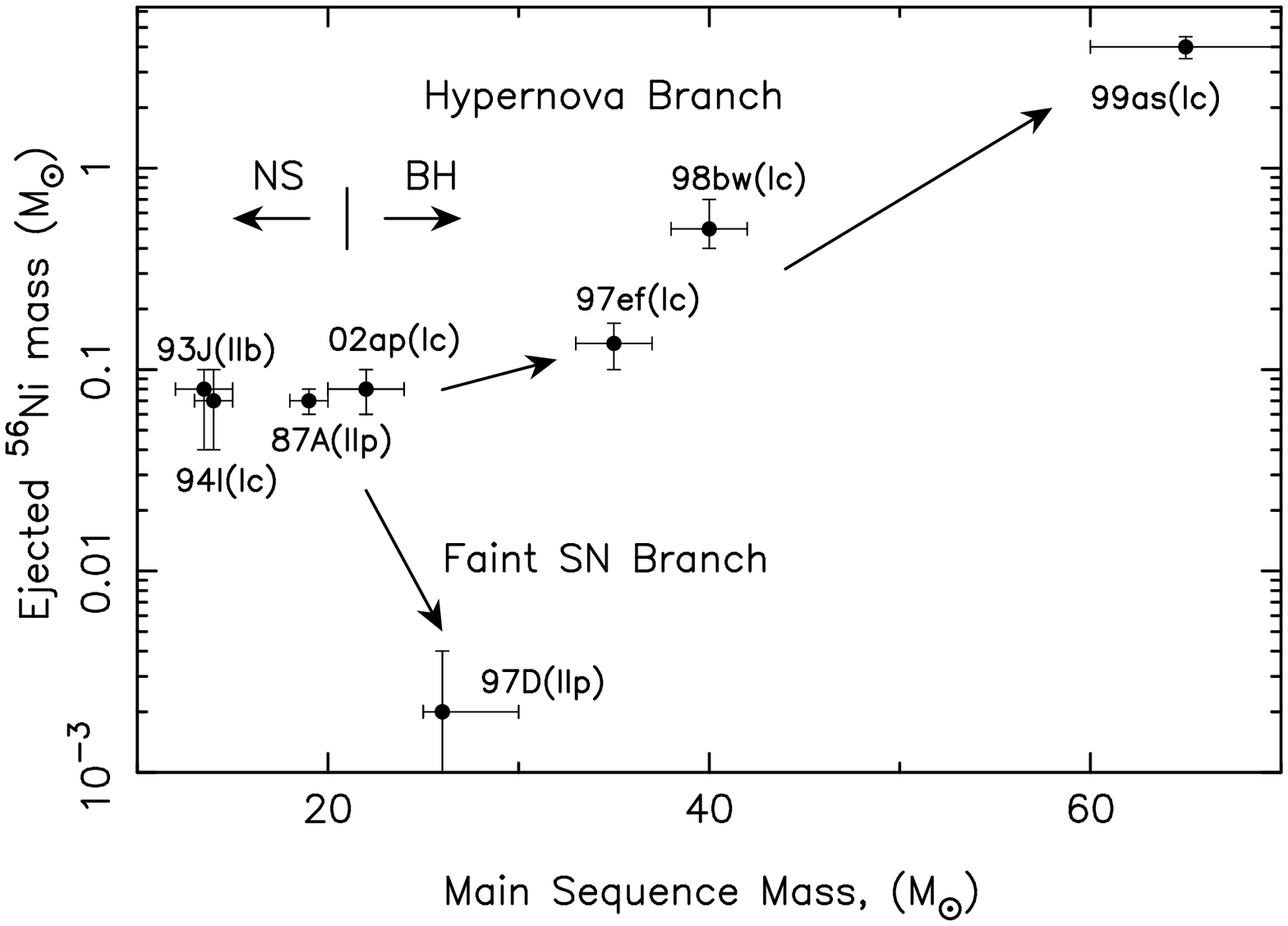}
\caption{The explosion energy and the ejected $^{56}$Ni mass as a
function of the main sequence mass of the progenitors for several
supernovae/hypernovae (Nomoto et al. 2003).}
\label{fig2}
\end{figure}

In contrast, SNe II 1997D and 1999br were very faint SNe with very low
KE (Turatto et al. 1998; Hamuy 2003; Zampieri et al. 2003).  In the
diagram that shows $E$ and the mass of $^{56}$Ni ejected $M(^{56}$Ni)
as a function of the main-sequence mass $M_{\rm ms}$ of the progenitor
star (Figure~\ref{fig2}), therefore, we propose that SNe from stars
with $M_{\rm ms} \mathrel{\rlap{\lower 4pt \hbox{\hskip 1pt
$\sim$}}\raise 1pt \hbox {$>$}}$ 20-25 $M_\odot$ have different $E$
and $M(^{56}$Ni), with a bright, energetic ``hypernova branch'' at one
extreme and a faint, low-energy SN branch at the other (Nomoto et
al. 2003).  For the faint SNe, the explosion energy was so small that
most $^{56}$Ni fell back onto the compact remnant.  Thus the faint SN
branch may become a ``failed'' SN branch at larger $M_{\rm ms}$.
Between the two branches, there may be a variety of SNe (Hamuy 2003).

This trend might be interpreted as follows.  Stars with $M_{\rm ms}
\mathrel{\rlap{\lower 4pt \hbox{\hskip 1pt $\sim$}}\raise 1pt \hbox
{$<$}}$ 20-25 $M_\odot$ form a neutron star, producing $\sim$ 0.08
$\pm$ 0.03 $M_\odot$ $^{56}$Ni as in SNe 1993J, 1994I, and 1987A.
Stars with $M_{\rm ms} \mathrel{\rlap{\lower 4pt \hbox{\hskip 1pt
$\sim$}}\raise 1pt \hbox {$>$}}$ 20-25 $M_\odot$ form a black hole;
whether they become hypernovae or faint SNe may depend on the angular
momentum in the collapsing core, which in turn depends on the stellar
winds, metallicity, magnetic fields, and binarity.  Hypernovae might
have rapidly rotating cores owing possibly to the spiraling-in of a
companion star in a binary system.

\begin{figure}[!ht]
\plotfiddle{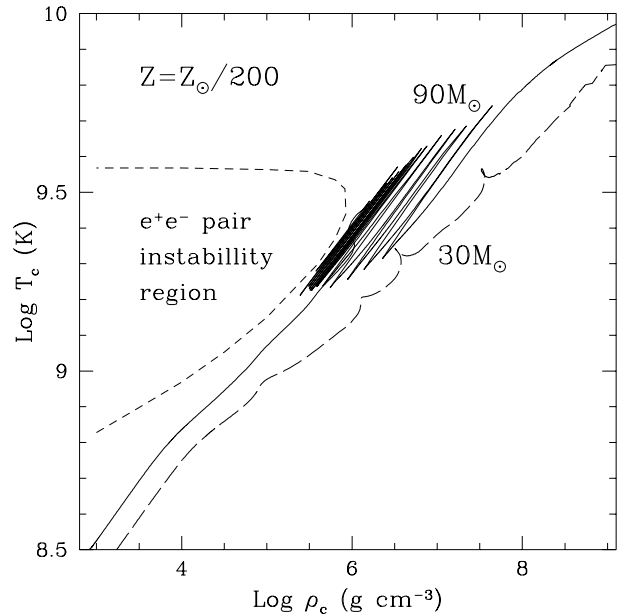}{2.9in}{0.}{42.}{42.}{-135.}{-65.}
\caption{Evolution of the central density and temperature for the 30
and 90 $M_\odot$ models (Umeda \& Nomoto 2004)
\label{rhot}}
\end{figure}

\section{Synthesis of $^{56}$Ni in $\sim 100 M_\odot$ Stars}

The light curve modeling of the unusually bright hypernova SN1999as
suggests that the progenitor is a core-collapse supernova and the
ejected $^{56}$Ni mass is as large as $\sim 3-4 M_\odot$.  Motivated
by SN 1990as, Umeda \& Nomoto (2004) have investigated how
much $^{56}$Ni can be synthesized in core-collapse massive supernovae.

 The evolutions of several very massive stars with initial masses of
$M \leq 100 M_\odot$ and low metallicity ($Z=Z_\odot/200$) have been
calculated from the main-sequence to ``hypernova'' explosions.  The
synthesized $^{56}$Ni mass increases with the increasing explosion
energy and the progenitor mass.  Umeda \& Nomoto (2004) found that for
the explosion energy of 3$\times 10^{52}$ ergs, for example, the
$^{56}$Ni mass of up to 2.2, 2.3, 5.0, and 6.6 $M_\odot$ can be
produced for the progenitors with masses of 30, 50, 80 and 100
$M_\odot$, that are sufficiently large to explain SN 1999as.

 Figure \ref{rhot} shows the evolution of the central density and
temperature for the 30 and 90$M_\odot$ models.  More massive stars
have larger specific entropy at the center, thus having higher
temperature for the same density.  For 90$M_\odot$, the evolutinary
track is very close to (but outside of) the ``e$^-$e$^+$
pair-instabillity region'' of $\Gamma < 4/3$ where $\Gamma$ denotes
the adiabatic index.  The evolution of the central temperature and
density is significantly different between the 30 and 90$M_\odot$
models during Si-burning at $T_9 = T/10^9$K =$ 2.5-4$.  The central
temperature and density of the 90$M_\odot$ model oscillate several
times. This is because in such massive stars radiation pressure is so
dominant that $\Gamma$ is close to 4/3, and thus the inner core of the
stars easily expands with the nuclear energy released by Si-burning.
Once it expands, the temperature drops suddenly, the central
Si-burning stops, and the stellar core turns into shrink.  Since only
small amount of Si is burnt for each cycle, this pulsations occur many
times.

 Umeda \& Nomoto (2004) found from the study of 80$ - 100 M_\odot$
stars that the number of the oscillations depends on the convective
parameter $f_k$: larger $f_k$ increases the number of the oscillation.
This is because for a larger $f_k$ fresh Si is mixed more efficiently
into the center, which increase the lifetime of this stage. The
situation is similar to the breathing pulse phase at the end of
He-burning, while the variation of the temperature and density is much
larger in Si burning than in the breathing pulse phase.  The amplitude
of the temperature and density variation is larger for more massive
stars, which suggests more and more drastic oscillations would occur
for larger mass stars.

\section{Nucleosynthesis in Hypernovae}

In core-collapse supernovae/hypernovae, stellar material undergoes
shock heating and subsequent explosive nucleosynthesis. Iron-peak
elements are produced in two distinct regions, which are characterized
by the peak temperature, $T_{\rm peak}$, of the shocked material.  For
$T_{\rm peak} > 5\times 10^9$K, material undergoes complete Si burning
whose products include Co, Zn, V, and some Cr after radioactive
decays.  For $4\times 10^9$K $<T_{\rm peak} < 5\times 10^9$K,
incomplete Si burning takes place and its after decay products include
Cr and Mn (Nakamura et al. 1999).

\subsection {Supernovae vs. Hypernovae}

\begin{figure}[!ht]
\plottwo{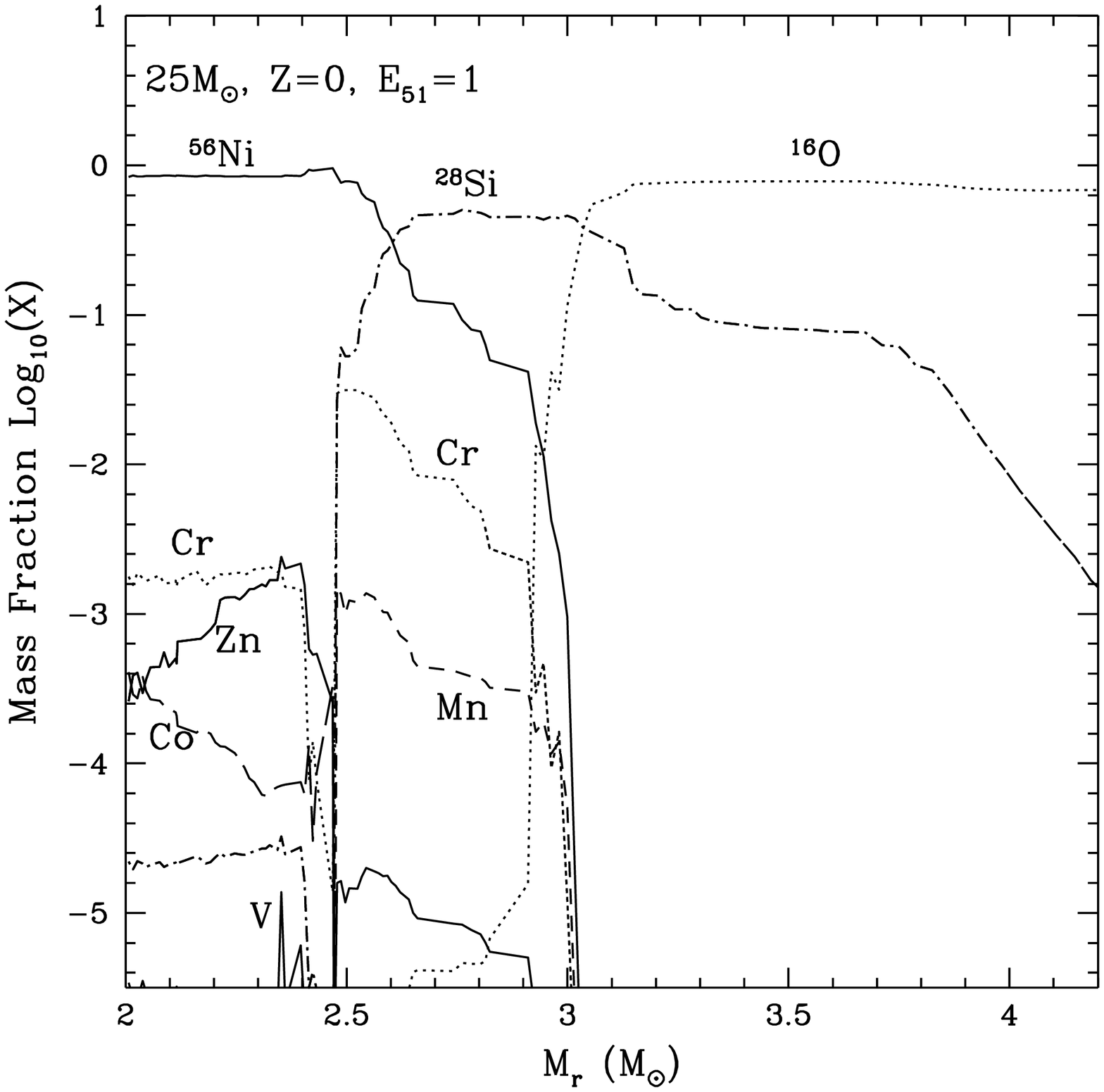}{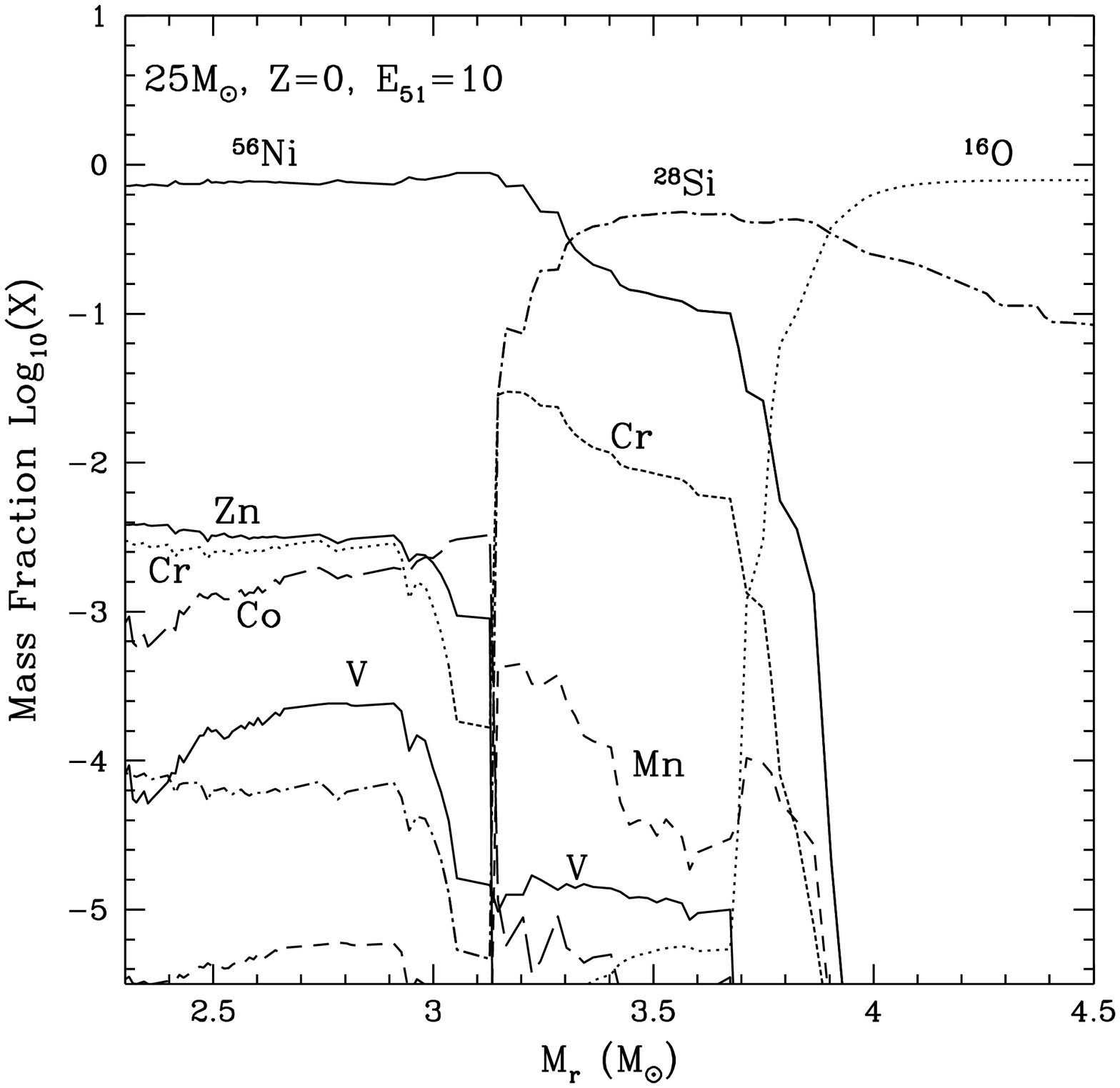}
\caption{Abundance distribution against the enclosed mass $M_r$ after
the explosion of Pop III 25 $M_\odot$ stars with $E_{51} = 1$ (left)
and $E_{51} = 10$ (right) (Umeda \& Nomoto 2002). }
\label{fig3}
\end{figure}

The right panel of Figure~\ref{fig3} shows the composition in the
ejecta of a 25 $M_\odot$ hypernova model ($E_{51} = 10$).  The
nucleosynthesis in a normal 25 $M_\odot$ SN model ($E_{51} = 1$) is
also shown for comparison in the left panel of Figure~\ref{fig3}
(Umeda \& Nomoto 2002).

We note the following characteristics of nucleosynthesis with very
large explosion energies (Nakamura et al. 2001b; Nomoto et al. 2001;
Umeda \& Nomoto 2005):

(1) Both complete and incomplete Si-burning regions shift outward in
mass compared with normal supernovae, so that the mass ratio between
the complete and incomplete Si-burning regions becomes larger.  As a
result, higher energy explosions tend to produce larger [(Zn, Co,
V)/Fe] and smaller [(Mn, Cr)/Fe], which can explain the trend observed
in very metal-poor stars (Umeda \& Nomoto 2005).

(2) In the complete Si-burning region of hypernovae, elements produced
by $\alpha$-rich freezeout are enhanced.  Hence, elements synthesized
through capturing of $\alpha$-particles, such as $^{44}$Ti, $^{48}$Cr,
and $^{64}$Ge (decaying into $^{44}$Ca, $^{48}$Ti, and $^{64}$Zn,
respectively) are more abundant.

(3) Oxygen burning takes place in more extended regions for the larger
KE.  Then more O, C, Al are burned to produce a larger amount of
burning products such as Si, S, and Ar.  Therefore, hypernova
nucleosynthesis is characterized by large abundance ratios of
[Si,S/O], which can explain the abundance feature of M82 (Umeda et
al. 2002).

\subsection{Hypernovae and Zn, Co, Mn, Cr}

Hypernova nucleosynthesis may have made an important contribution to
Galactic chemical evolution.  In the early galactic epoch when the
galaxy was not yet chemically well-mixed, [Fe/H] may well be
determined by mostly a single SN event (Audouze \& Silk 1995). The
formation of metal-poor stars is supposed to be driven by a supernova
shock, so that [Fe/H] is determined by the ejected Fe mass and the
amount of circumstellar hydrogen swept-up by the shock wave (Ryan,
Norris, \& Beers 1996).  Then, hypernovae with larger $E$ are likely
to induce the formation of stars with smaller [Fe/H], because the mass
of interstellar hydrogen swept up by a hypernova is roughly
proportional to $E$ (Ryan et al. 1996; Shigeyama \& Tsujimoto 1998)
and the ratio of the ejected iron mass to $E$ is smaller for
hypernovae than for normal supernovae.

\begin{figure}[!ht]
\plotfiddle{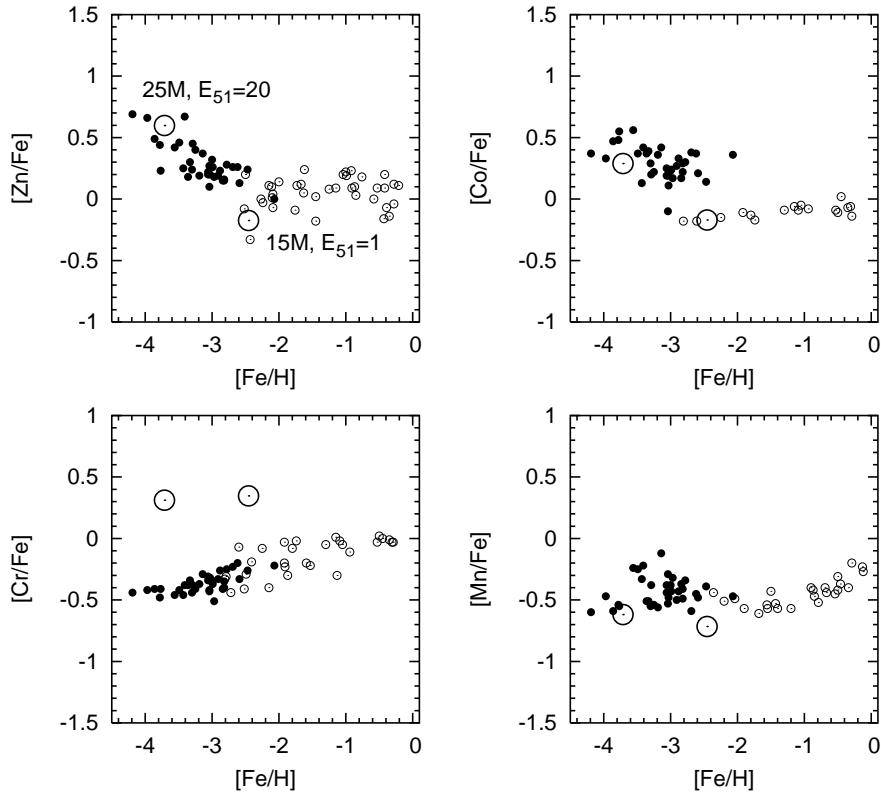}{3.9in}{0.}{120.}{120.}{-230.}{-60.}
\caption{Observed abundance ratios of [Zn, Co, Cr, Mn/Fe] vs [Fe/H]
(Cayrel et al. 2004) compared with (15 $M_\odot$, $E_{51}=1$) and (25
$M_\odot$, $E_{51}$=30) models (Tominaga et al. 2005).}
\label{fig4}
\end{figure}

In the observed abundances of halo stars, there are significant
differences between the abundance patterns in the iron-peak elements
below and above [Fe/H]$ \sim -2.5$ - $-3$.

(1) For [Fe/H]$\mathrel{\rlap{\lower 4pt \hbox{\hskip 1pt
$\sim$}}\raise 1pt \hbox {$<$}} -2.5$, the mean values of [Cr/Fe] and
[Mn/Fe] decrease toward smaller metallicity, while [Co/Fe] increases
(McWilliam et al. 1995; Ryan et al. 1996).

(2) [Zn/Fe]$ \sim 0$ for [Fe/H] $\simeq -3$ to $0$ (Sneden, Gratton,
\& Crocker 1991), while at [Fe/H] $< -3.3$, [Zn/Fe] increases toward
smaller metallicity (Cayrel et al. 2004).

The larger [(Zn, Co)/Fe] and smaller [(Mn, Cr)/Fe] in the supernova
ejecta can be realized if the mass ratio between the complete Si
burning region and the incomplete Si burning region is larger, or
equivalently if deep material from the complete Si-burning region is
ejected by mixing or aspherical effects.  This can be realized if (1)
the mass cut between the ejecta and the compact remnant is located at
smaller $M_r$ (Nakamura et al. 1999), (2) $E$ is larger to move the
outer edge of the complete Si burning region to larger $M_r$ (Nakamura
et al. 2001b), or (3) asphericity in the explosion is larger.

Among these possibilities, a large explosion energy $E$ enhances
$\alpha$-rich freezeout, which results in an increase of the local
mass fractions of Zn and Co, while Cr and Mn are not enhanced (Umeda
\& Nomoto 2002, 2005).  Models with $E_{51} = 1 $ do not produce
sufficiently large [Zn/Fe].  To be compatible with the observations of
[Zn/Fe] $\sim 0.5$, the explosion energy must be much larger, i.e.,
$E_{51} \mathrel{\rlap{\lower 4pt \hbox{\hskip 1pt $\sim$}}\raise 1pt
\hbox {$>$}} 20$ for $M \mathrel{\rlap{\lower 4pt \hbox{\hskip 1pt
$\sim$}}\raise 1pt \hbox {$>$}} 20 M_\odot$, i.e., hypernova-like
explosions of massive stars ($M \mathrel{\rlap{\lower 4pt \hbox{\hskip
1pt $\sim$}}\raise 1pt \hbox {$>$}} 25 M_\odot$) with $E_{51} > 10$
are responsible for the production of Zn.

In the hypernova models, the overproduction of Ni, as found in the
simple ``deep'' mass-cut model, can be avoided (Umeda \& Nomoto 2005).
Therefore, if hypernovae made significant contributions to the early
Galactic chemical evolution, it could explain the large Zn and Co
abundances and the small Mn and Cr abundances observed in very
metal-poor stars (Fig.~\ref{fig4}: Tominaga et al. 2005).

\begin{figure}[!ht]
\plottwo{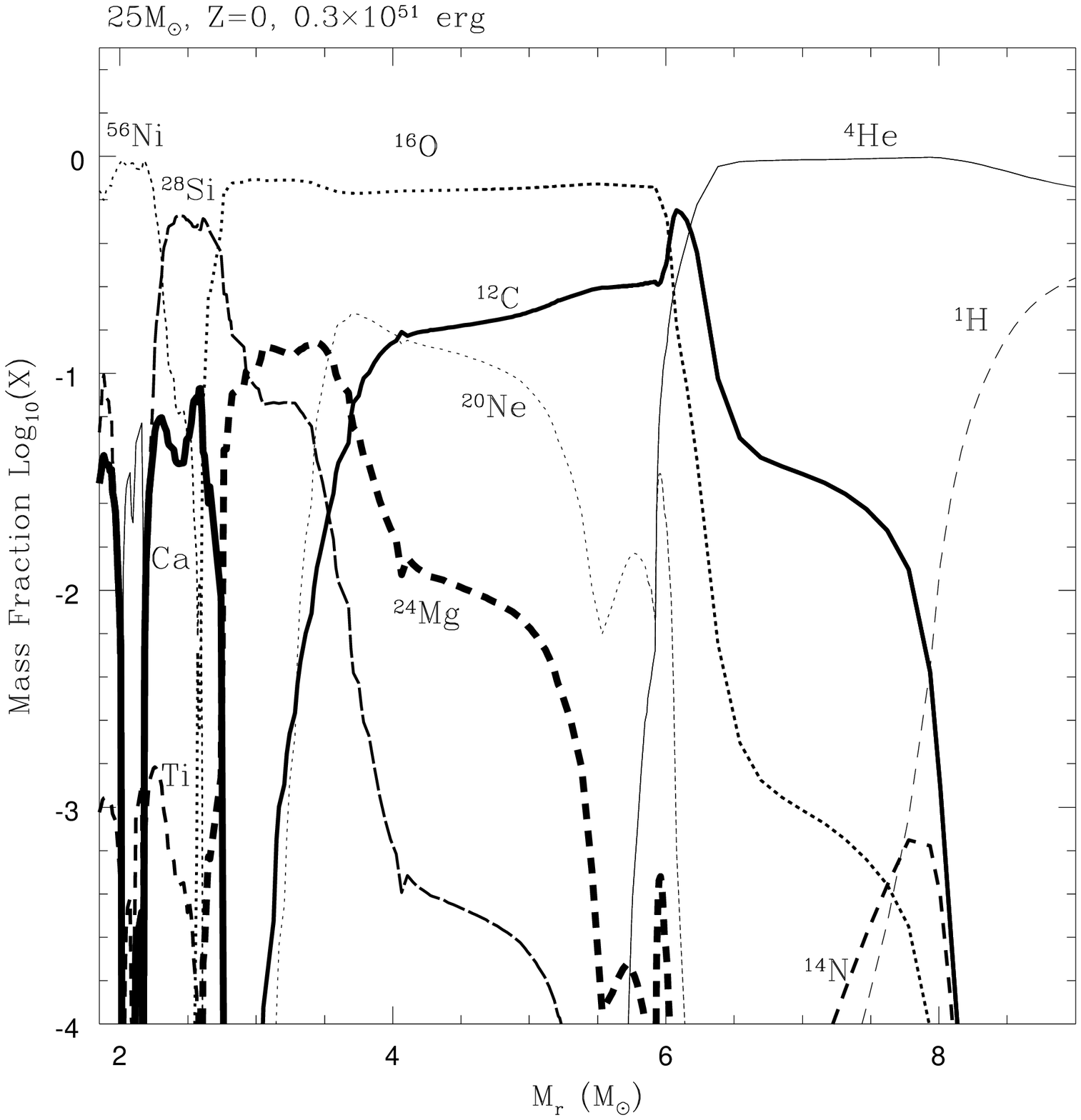}{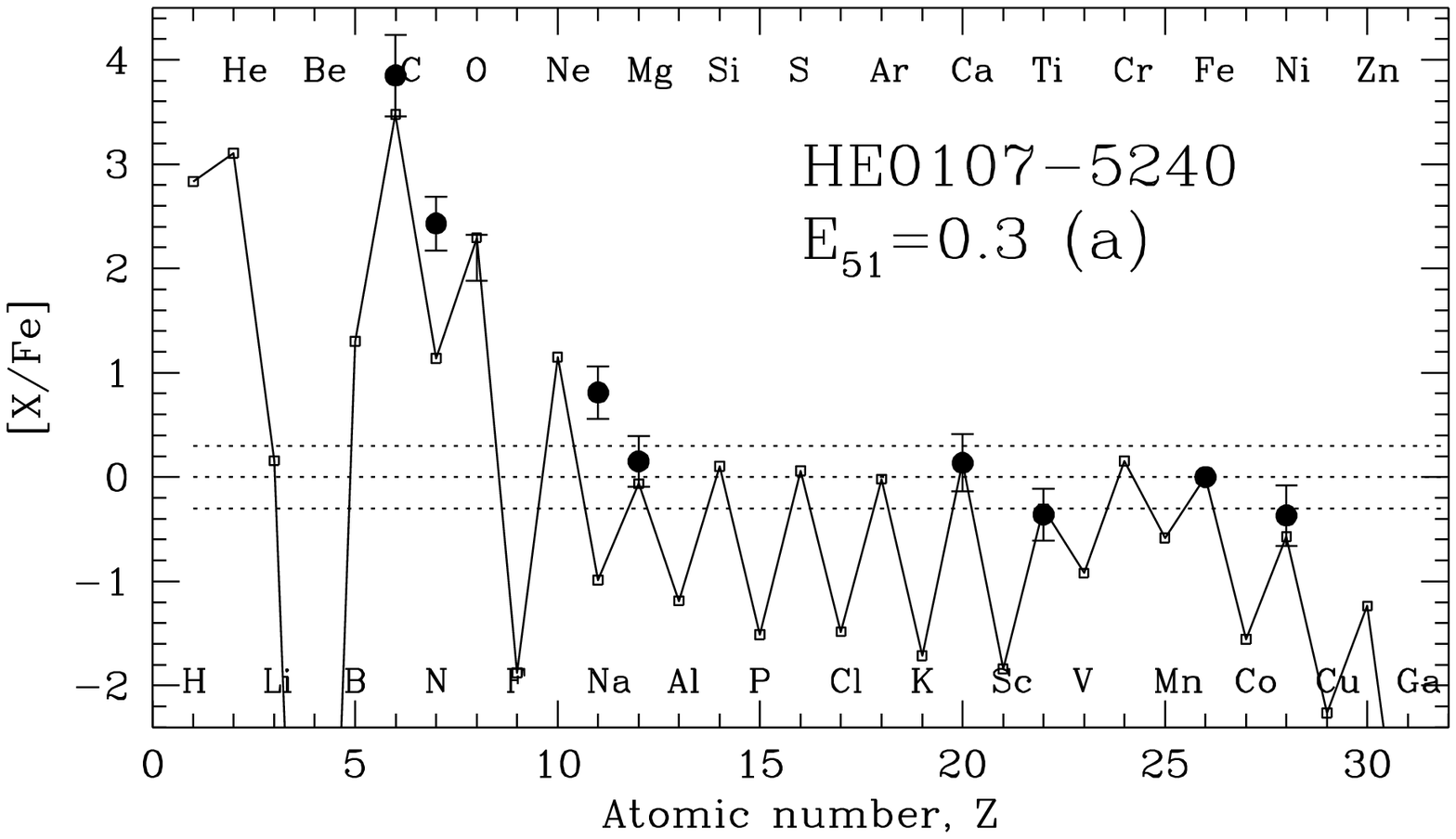}
\caption{(left) The post-explosion abundance distributions for the 25
$M_\odot$ model with the explosion energy $E_{51} =$ 0.3.  (right)
Elemental abundances of the C-rich, most Fe deficient star HE0107-5240
(Christlieb et al. 2004: filled circles), compared with a theoretical
supernova yield (Umeda \& Nomoto 2003, 2005).}
\label{fig5}
\end{figure}

\section{Extremely Metal-Poor (EMP) Stars}

Recently the most Fe deficient and C-rich low mass star, HE0107-5240,
was discovered (Christlieb et al. 2002, 2004).  This star has [Fe/H]
$= - 5.3$ but its mass is as low as 0.8 $M_\odot$.  This would
challenge the recent theoretical arguments that the formation of low
mass stars, which should survive until today, is suppressed below
[Fe/H] $= -4$ (Schneider et al. 2002).

The important clue to this problem is the observed abundance pattern
of this star.  This star is characterized by a very large ratios of
[C/Fe] = 4.0 and [N/Fe] = 2.3, while the abundances of elements
heavier than Mg are as low as Fe (Christlieb et al. 2002).
Interestingly, this is not the only extremely metal poor (EMP) stars
that have the large C/Fe and N/Fe ratios, but several other such stars
have been discovered (Aoki et al. 2002).  Therefore the reasonable
explanation of the abundance pattern should explain other EMP stars as
well.  We show that the abundance pattern of C-rich EMP stars can be
reasonably explained by the nucleosynthesis of 20 - 130 $M_\odot$
supernovae with various explosion energies and the degree of mixing
and fallback of the ejecta.

\subsection{The Most Fe-Poor Star HE0107-5240}

We consider a model that C-rich EMP stars are produced in the ejecta
of (almost) metal-free supernova mixed with extremely metal-poor
interstellar matter.  We use Pop III pre-supernova progenitor models,
simulate the supernova explosion and calculate detailed
nucleosynthesis (Umeda \& Nomoto 2003).

In Figure~\ref{fig5} (right) we show that the elemental abundances of
one of our models are in good agreement with HE0107-5240, where the
progenitor mass is 25 $M_\odot$ and the explosion energy $E_{51} =$ 0.3
(Umeda \& Nomoto 2003).

In this model, explosive nucleosynthesis takes place behind the shock
wave that is generated at $M_r =$ 1.8 $M_\odot$ and propagates
outward. The resultant abundance distribution is seen in
Figure~\ref{fig5} (left), where $M_r$ denotes the Lagrangian mass
coordinate measured from the center of the pre-supernova model.  The
processed material is assumed to mix uniformly in the region from $M_r
=$ 1.8 $M_\odot$ and 6.0 $M_\odot$.  Such a large scale mixing was
found to take place in SN1987A and various explosion models (Hachisu
et al. 1990).  Almost all materials below $M_r =$ 6.0 $M_\odot$ fall
back to the central remnant and only a small fraction ($f = 2 \times$
10$^{-5}$) is ejected from this region.  The ejected Fe mass is 8
$\times$ 10$^{-6}$ $M_\odot$.

The CNO elements in the ejecta were produced by pre-collapse He shell
burning in the He-layer, which contains 0.2 $M_\odot$ $^{12}$C.
Mixing of H into the He shell-burning region produces 4 $\times$
10$^{-4}$ $M_\odot$ $^{14}$N.  On the other hand, only a small amount
of heavier elements (Mg, Ca, and Fe-peak elements) are ejected and
their abundance ratios are the average in the mixed region.  The
sub-solar ratios of [Ti/Fe] $= -0.4$ and [Ni/Fe] $= -0.4$ are the
results of the relatively small explosion energy ($E_{51} =$ 0.3).
With this "mixing and fallback", the large C/Fe and C/Mg ratios
observed in HE0107-5240 are well reproduced.

In this model, N/Fe appears to be underproduced. However, N can be
produced inside the EMP stars through the C-N cycle, and brought up to
the surface during the first dredge up stage while becoming a
red-giant star (Boothroyd \& Sackmann 1999).
 
\begin{figure}[!ht]
\plottwo{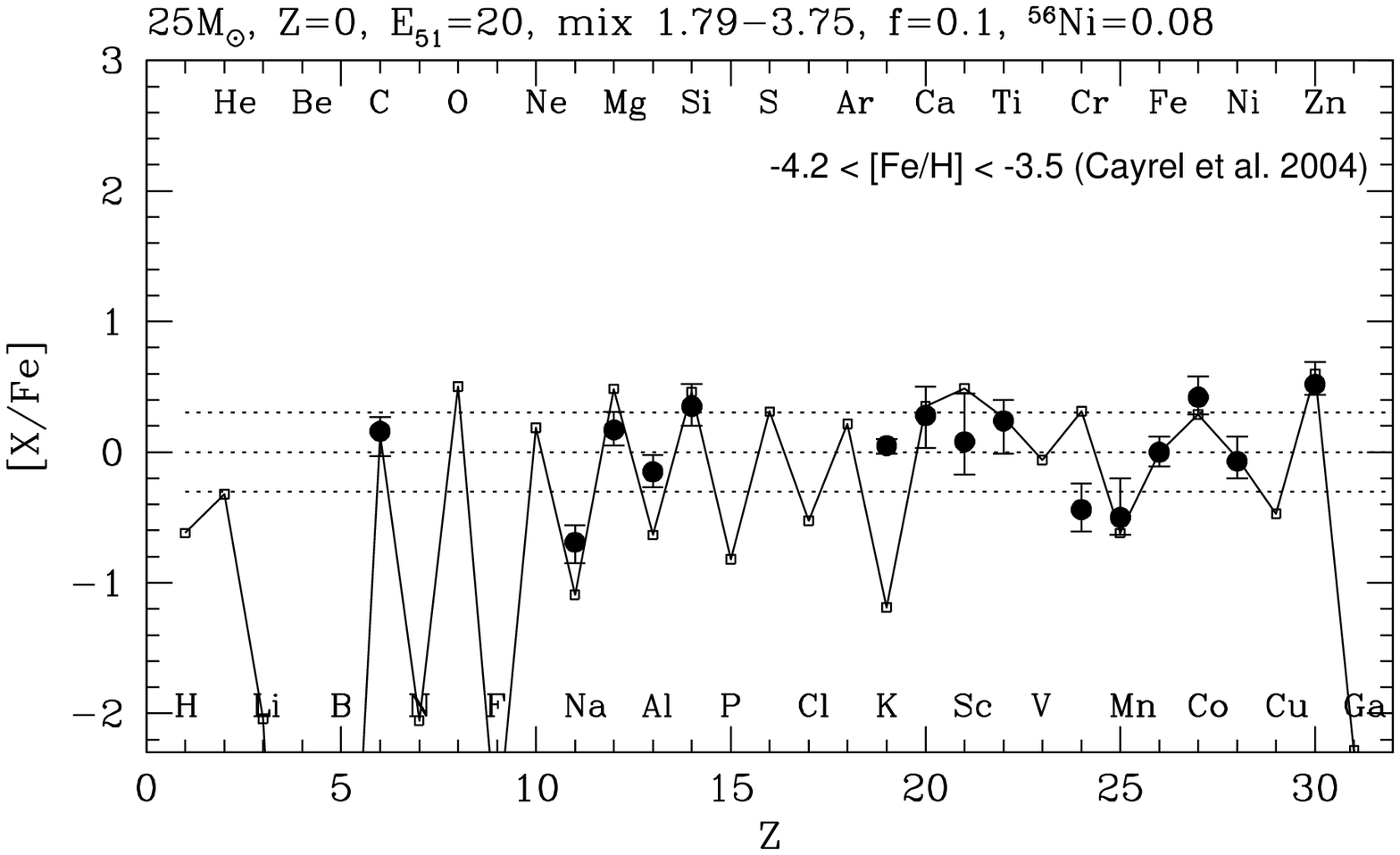}{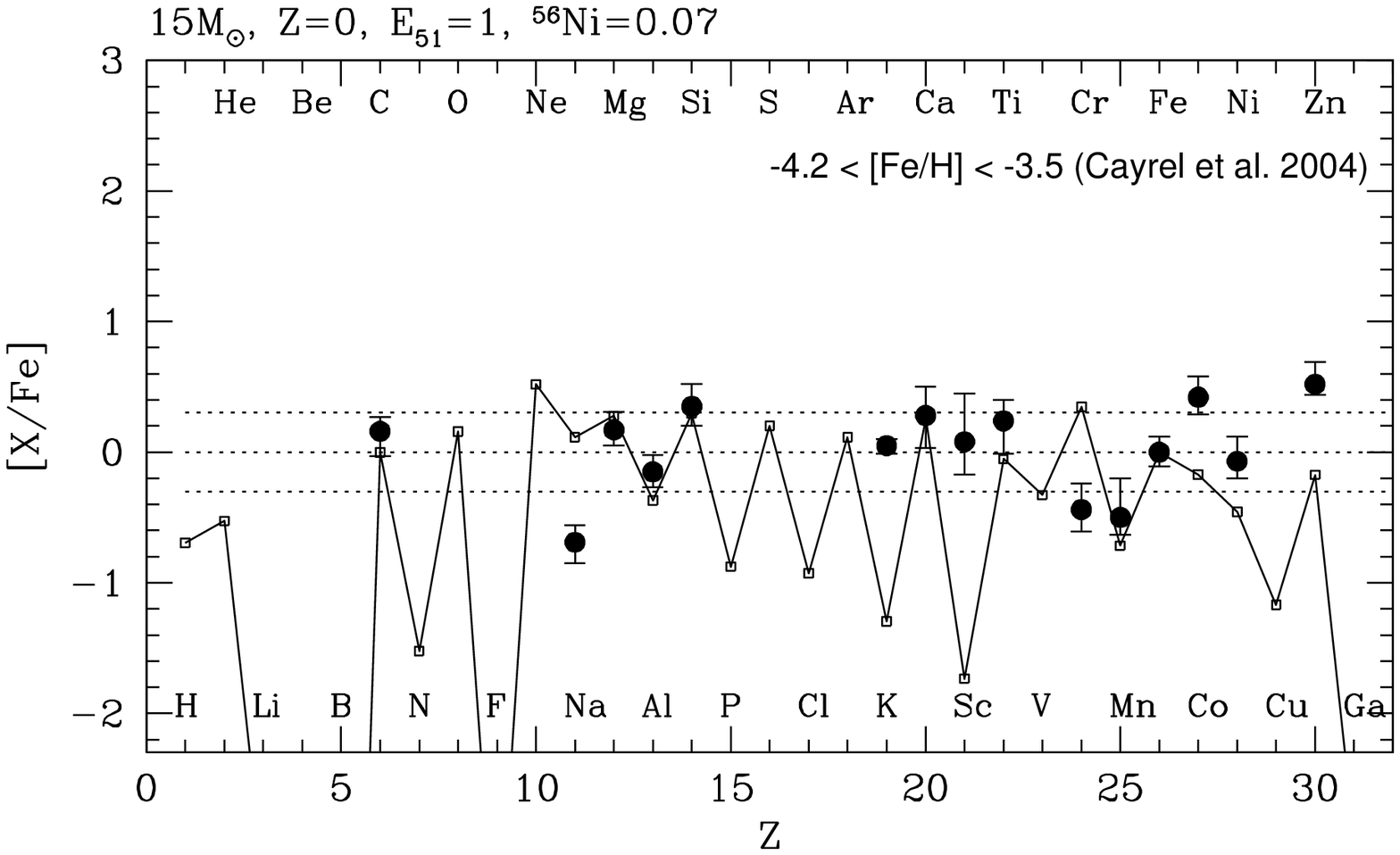}
\caption{Averaged elemental abundances of stars with [Fe/H] $= -3.7$
(Cayrel et al. 2004) compared with the hypernova yield (left: 25
$M_\odot$, $E_{51} =$ 20), and the normal SN yield (right: 15
$M_\odot$, $E_{51} =$ 1).}
\label{fig7}
\end{figure}

\subsection{EMP Stars from VLT Observations}

The "mixing and fall back" process can reproduce the abundance pattern
of the typical EMP stars without enhancement of C and N.
Figure~\ref{fig7} shows that the averaged abundances of [Fe/H] $=
-3.7$ stars in Cayrel et al. (2003) can be fitted well with the
hypernova model of 25 $M_\odot$ and $E_{51} =$ 20 (left) but not with
the normal SN model of 15 $M_\odot$ and $E_{51} =$ 1 (right) (Tominaga
et al. 2005).

\section{First Stars}

It is of vital importance to identify the first generation stars in
the Universe, i.e., totally metal-free, Pop III stars.  The impact of
the formation of Pop III stars on the evolution of the Universe
depends on their typical masses.

\subsection{High Mass vs. Low Mass}

Recent numerical models have shown that, the first stars are as
massive as $\sim$ 100 $M_\odot$ (Abel et al. 2002).  The formation of
long-lived low mass Pop III stars may be inefficient because of slow
cooling of metal free gas cloud, which is consistent with the failure
of attempts to find Pop III stars.

If the most Fe deficient star, HE0107-5240, is a Pop III low mass star
that has gained its metal from a companion star or interstellar matter
(Yoshii 1981), would it mean that the above theoretical arguments
are incorrect and that such low mass Pop III stars have not been
discovered only because of the difficulty in the observations?

Based on the results in the earlier section, we propose that the first
generation supernovae were the explosion of $\sim$ 20-130 $M_\odot$
stars and some of them produced C-rich, Fe-poor ejecta.  Then the low
mass star with even [Fe/H] $< -5$ can form from the gases of mixture
of such a supernova ejecta and the (almost) metal-free interstellar
matter, because the gases can be efficiently cooled by enhanced C and
O ([C/H] $\sim -1$).

\subsection{Pair Instability SNe vs. Core Collapse SNe}

In contrast to the core-collapse supernovae of 20-130 $M_\odot$ stars,
the observed abundance patterns cannot be explained by the explosions
of more massive, 130 - 300 $M_\odot$ stars. These stars undergo
pair-instability supernovae (PISNe) and are disrupted completely
(e.g., Umeda \& Nomoto 2002; Heger \& Woosley 2002), which cannot be
consistent with the large C/Fe observed in HE0107-5240 and other
C-rich EMP stars.  The abundance ratios of iron-peak elements ([Zn/Fe]
$< -0.8$ and [Co/Fe] $< -0.2$) in the PISN ejecta (Fig.~\ref{fig7};
Umeda \& Nomoto 2002; Heger \& Woosley 2002) cannot explain the large
Zn/Fe and Co/Fe in the typical EMP stars (McWilliam et al. 1995;
Norris et al. 2001; Cayrel et al. 2003) and CS22949-037 either.
Therefore the supernova progenitors that are responsible for the
formation of EMP stars are most likely in the range of $M \sim 20 -
130$ $M_\odot$, but not more massive than 130 $M_\odot$.  This upper
limit depends on the stability of massive stars as will be discussed
below.

\begin{table}[t]
\caption{The results of the stability analysis for Pop III and Pop I
stars.  $\bigcirc$ and $\times$ represent that the star is stable and
unstable, respectively.  The $e$-folding time for the fundamental mode
is shown after $\times$ in units of $10^4$yr (Nomoto et al. 2003).}
\begin{center}
\footnotesize
\begin{tabular}{ccccccc}
\hline \hline
{\large mass ($M_\odot$)} &{\large 80}&{\large 100}&{\large 
120}&{\large 150} &{\large 180} &{\large 300} \\ \hline
{\large Pop III} &{\large $\bigcirc$ }&{\large $\bigcirc$ }&{\large
$\bigcirc$ }&{\large $\times$ (9.03)} &{\large 
 $\times$ (4.83)} &{\large $\times$ (2.15)} \\ 
{\large Pop I }&{\large $\bigcirc$ }&{\large $\times$ (7.02)} &{\large
 $\times$ (2.35)} &{\large $\times$ (1.43)} &{\large $\times$ (1.21)}
 &{\large $\times$ (1.71)} \\ \hline
\end{tabular}
\end{center}
\label{tab:pop3}
\end{table}

\subsection{Stability and Mass Loss of Massive Pop III Stars}

To determine the upper limit mass of the Zero Age Main Sequence
(ZAMS), Nomoto et al. (2003) analyzed a linear non-adiabatic stability
of massive ($80M_{\odot}$ - $300M_{\odot}$) Pop III stars using a
radial pulsation code. Because CNO elements are absent during the
early stage of their evolution, the CNO cycle does not operate and the
star contracts until temperature rises sufficiently high for the
$3\alpha$ reaction to produce $^{12}$C.  We calculate that these stars
have $X_{\rm CNO} \sim 1.6 - 4.0\times10^{-10}$, and the central
temperature $T_{c}\sim1.4\times10^8 K$ on their ZAMS.  We also examine
the models of Pop I stars for comparison.

Table~\ref{tab:pop3} shows the results for our analysis. The critical
mass of ZAMS Pop III star is $128M_{\odot}$ while that of Pop I star
is $94M_{\odot}$.  This difference comes from very compact structures
(with high $T_c$) of Pop III stars.

Stars more massive than the critical mass will undergo pulsation and
mass loss. We note that the $e$-folding time of instability is much
longer for Pop III stars than Pop I stars with the same mass, and thus
the mass loss rate is much lower. These results are consistent with
Ibrahim et al. (1981) and Baraffe et al. (2001).  However, the absence
of the indication of PISNe in EMP stars might imply that these massive
stars above 130$M_{\odot}$ undergo significant mass loss, thus
evolving into Fe core-collapse rather than PISNe.

\section{Type Ia/IIn Supernovae: SNe~2002ic, 1997cy, and 1999E}

\begin{figure}[!ht]
\plotfiddle{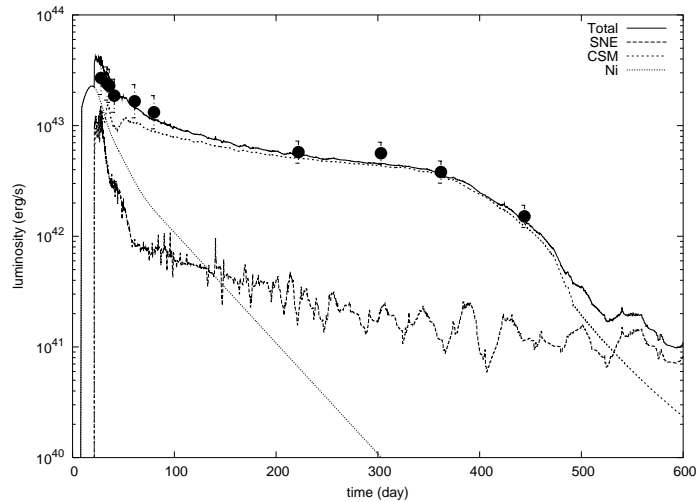}{2.2in}{0.}{75.}{75.}{-178.}{-35.}
\caption{Model light curve (thick line: Nomoto et al. 2005)
compared with the observation of SN 2002ic (filled circles; Deng
et al. 2004).
\label{sn02ic}}
\end{figure}

     SNe~1997cy and 1999E were initially classified as Type IIn
because they showed H$\alpha$ emission. SN~2002ic would also have been
so classified, had it not been discovered at an early epoch.
SN~1997cy ($z=0.063$) is among the most luminous SNe discovered so far
($M_{V}<-20.1$ about maximum light), and SN~1999E is also bright
($M_{V}<-19.5$).  Both SNe~1997cy and 1999E have been suspected to be
spatially and temporally related to a GRB (Germany et al. 2000; Rigon
et al. 2003).  However, both the classification and the associations
with a GRB must now be seen as highly questionable in view of the fact
that their replica, SN~2002ic, appears to have been a genuine SN Ia at
an earlier phase (Hamuy et al. 2003; Deng et al. 2004).

     Nomoto et al. (2005) calculated the interaction between the
expanding ejecta and CSM (circumstellar matter).  For the supernova
ejecta, the carbon deflagration model W7 was used.  After day $\sim$
350, the light curve starts declining.  To reproduce the declining
part of the light curve, we add the outer CSM of 0.2 $M_\odot$ where
the density declines sharply as $n = 6$.  This implies that the total
mass of CSM is $\sim$ 1.3 $M_\odot$.

\end{document}